\documentclass[conference]{IEEEtran}
\IEEEoverridecommandlockouts

\usepackage{url}
\usepackage{cite}
\usepackage{booktabs}
\usepackage{amsmath,amssymb,amsfonts}
\usepackage{algorithmic}
\usepackage{graphicx}
\usepackage{textcomp}
\usepackage{xcolor}
\def\BibTeX{{\rm B\kern-.05em{\sc i\kern-.025em b}\kern-.08em
    T\kern-.1667em\lower.7ex\hbox{E}\kern-.125emX}}
\begin{document}

\title{\texttt{VeriCWEty}: Embedding enabled Line-Level CWE Detection in Verilog
}
\author{ 
\IEEEauthorblockN{Prithwish Basu Roy}
\IEEEauthorblockA{
\textit{NYU Tandon School of Engineering}\\
New York, US \\
pb2718@nyu.edu
}
\and
\IEEEauthorblockN{Zeng Wang}
\IEEEauthorblockA{\textit{NYU Abu Dhabi} \\
Abu Dhabi, UAE\\
zw3464@nyu.edu
}
\and
\IEEEauthorblockN{Anatolii Chuvashlov}
\IEEEauthorblockA{\textit{NYU Abu Dhabi} \\
Abu Dhabi, UAE\\
ac10756@nyu.edu
}
\and
\IEEEauthorblockN{Weihua Xiao}
\IEEEauthorblockA{\textit{NYU Tandon School of Engineering} \\
New York, US \\
weihua.xiao@nyu.edu
}
\and
\IEEEauthorblockN{Johann Knechtel}
\IEEEauthorblockA{\textit{NYU Abu Dhabi} \\
Center for Cyber Security \\
Abu Dhabi, UAE\\
johann@nyu.edu
}
\and
\IEEEauthorblockN{Ozgur Sinanoglu}
\IEEEauthorblockA{\textit{NYU Abu Dhabi} \\
Center for Cyber Security \\
Abu Dhabi, UAE\\
ozgursin@nyu.edu
}
\and
\IEEEauthorblockN{Ramesh Karri}
\IEEEauthorblockA{\textit{NYU Tandon School of Engineering} \\
New York, US \\
rkarri@nyu.edu
}}
\maketitle

\begin{abstract}
Large Language Models (LLMs) have shown significant improvement in RTL code generation. Despite the advances, the generated code is often riddled with common vulnerabilities and weaknesses (CWEs) that can slip by untrained eyes. Attackers can often exploit these weaknesses to fulfill their nefarious motives. Existing RTL bug-detection techniques rely on rule-based checks, formal properties, or coarse-grained structural analysis, which either fail to capture semantic vulnerabilities or lack precise localization. In our work, we bridge this gap by proposing an embedding-based bug-detection framework that detects and classifies bugs at both module and line-level granularity. Our method achieves about 89\% precision in identifying common CWEs such as CWE-1244 and CWE-1245, and 96\% accuracy in detecting line-level bugs. 
\end{abstract}


\section{Introduction}
Bug detection in modern hardware designs remains challenging, often requiring expertise in both hardware design and security analysis~\cite{pan2022survey, ahmad2023fixing}. As hardware systems grow in scale and complexity, manual inspection alone becomes insufficient for timely and comprehensive analysis. This limitation motivates scalable and automated vulnerability detection frameworks that improve verification efficiency, broaden bug coverage, and strengthen hardware security assurance~\cite{witharana2022survey,jayasena2024directed}.

Existing automated approaches fall short in different ways. Static analysis techniques~\cite{Bening2001} identify vulnerabilities by recognizing recurring patterns in code blocks or narrowing the search space through signals of interest~\cite{dontCWEAT}. While effective in constrained settings, these methods often lack the generalizability needed to scale across diverse Common Weakness Enumerations (CWEs), as they rely heavily on handcrafted rules or heuristics that cannot fully capture the semantic complexity of hardware vulnerabilities~\cite{ahmad2025lashed,long2025veriloglavd}. On the other hand, large language models have shown strong promise in software vulnerability detection and explanation~\cite{zhou2024large}. Recent studies have also begun to explore LLM-based representations for RTL design analysis and downstream hardware tasks~\cite{xiao2025trojanloc,hemadri2025veriloc}. However, translating these advances to hardware vulnerability detection remains non-trivial. First, hardware-security training data in HDLs such as Verilog is scarce, limiting model comprehension of hardware-specific CWEs and the micro-architectural contexts in which they arise~\cite{fu2023llm4sechw,wang2024llms,qayyum2025llm}. Second, AST-based and structural flow analyses, even when augmented with LLM-derived signals, generalize poorly across diverse vulnerability classes, as structural representations alone are insufficient to capture the contextual and semantic distinctions that differentiate hardware weaknesses~\cite{dontCWEAT, ahmad2025lashed}.

A fundamental limitation shared by these approaches is their implicit treatment of CWEs as a flat set of independent, mutually exclusive categories. In practice, CWEs are organized within a structured hierarchical taxonomy in which vulnerability classes are related through parent-child and sibling associations reflecting shared root causes, overlapping manifestation patterns, or common exploitation conditions~\cite{cwe_mitre_2026}. Accurate CWE annotation therefore demands not merely recognition of individual weakness patterns, but precise, context-aware discrimination among closely related entries within the hierarchy~\cite{rogers2024security}. Furthermore, the same CWE can manifest differently depending on the surrounding micro-architectural context, rendering pattern-matching and rule-based detection inherently brittle and difficult to generalize.

In this paper, we propose \texttt{VeriCWEty}, a vector embedding-assisted CWE detection framework for hardware vulnerabilities, the first of its kind to leverage the semantic power of LLM embeddings for this task. Rather than relying on syntactic patterns or handcrafted rules, \texttt{VeriCWEty} encodes the syntactic and semantic essence of each CWE into rich vector representations, enabling the framework to capture subtle distinctions between related CWEs and reason over design context. By operating at both the module level and the line level, \texttt{VeriCWEty} not only identifies the type of CWE present in a given hardware module, but also pinpoints the exact lines where the vulnerability occurs, offering fine-grained, actionable detection results. Our key contributions are as follows:

\begin{itemize}
    \item  We propose \texttt{VeriCWEty}, a framework that detects CWEs using vector embeddings.

    \item We propose a voting-based scheme using the state-of-the-art LLMs to automatically label buggy datasets.

    \item To the best of our knowledge, this is the first work that identifies the CWE and pinpoints its location in the code just based on vector embeddings.

    \item Our framework is able to detect crucial CWEs like CWE-1244 and CWE-1245 with a precision of 89\%.
\end{itemize}

\section{Background}
\subsection{CWE and its types}
Ensuring the development of bug-free code remains a persistent challenge for hardware designers. Developers frequently introduce similar defects across projects, often without understanding the underlying causes. When these vulnerabilities are later identified as critical, the cost of remediation increases substantially. To assist developers, the Common Weakness Enumeration (CWE) database, maintained by MITRE~\cite{cwe_mitre_2026}, catalogs frequently occurring software and hardware vulnerabilities. This resource provides detailed descriptions and potential solutions for each weakness. Each CWE is assigned a unique identifier; however, many CWEs are hierarchically related as parents or children and are not entirely independent. 

\begin{table}[t]
\centering
\scriptsize
\setlength{\tabcolsep}{4pt}
\renewcommand{\arraystretch}{1.1}
\caption{List of CWEs commonly used in the literature.}
\begin{tabular}{p{2cm} p{6cm}}
\hline
\textbf{CWE} & \textbf{Description} \\
\hline
CWE-250 & Execution with unnecessary privileges, increasing attack impact. \\
CWE-269 & Improper privilege management allowing unauthorized privilege changes. \\
CWE-284 & Improper access control enabling unauthorized resource access. \\
CWE-310 & Cryptographic issues due to weak or improper algorithms/usage. \\
CWE-321 & Use of hard-coded cryptographic key in code. \\
CWE-506 & Embedded malicious code (e.g., backdoor, trojan). \\
CWE-1191 & On-chip debug/test interface misuse leading to unauthorized access. \\
CWE-1231 & Improper hardware lock or protection mechanism implementation. \\
CWE-1233 & Security-sensitive hardware control exposed to untrusted agents. \\
CWE-1244 & Internal asset (e.g., key/data) exposed to unauthorized entity. \\
CWE-1245 & Shared resource allows unintended information leakage. \\
CWE-1260 & Improper isolation of hardware resources or components. \\
CWE-1300 & Improper protection of physical side-channel information. \\
\hline
\end{tabular}
\label{tab:cwe_summary}
\end{table}

\subsection{Why CWE annotations are difficult?} The classification of vulnerabilities into appropriate CWEs present significant challenges due to the non-independence of categories and their frequent hierarchical and overlapping relationships. For instance, CWE-284 (Improper Access Control) encompasses CWE-269 (Improper Privilege Management), which in turn relates to CWE-250 (Execution with Unnecessary Privileges). Likewise, CWE-321 (Hard-coded Cryptographic Key) represents a specific case of CWE-310 (Cryptographic Issues). In the hardware context, complexity increases, as weaknesses such as CWE-1260 (Improper Isolation) may lead to multiple subsequent issues, including CWE-1244 (Internal Asset Exposure) or CWE-1245 (Shared Resource Leakage), thereby establishing causal rather than strictly hierarchical connections. Furthermore, several CWEs (e.g., CWE-1191, CWE-1231, CWE-1233) serve as sibling categories that share semantic similarities but lack direct parent–child relationships. This combination of hierarchical structure, overlap, and causality complicates the application of machine learning models, which cannot treat CWEs as mutually exclusive labels and often necessitate hierarchical or multi-label classification strategies for accurate prediction.

\subsection{Methods for detecting CWEs} Vulnerability analysis and detection are complex tasks that typically require significant expertise~\cite{bidmeshki2021, hardfails2019}. Common approaches include static analysis (using linter ~\cite{Bening2001}), formal analysis~\cite{cheri,Ray2019}, and simulation-based analysis~\cite{rtlConTest}. Recently, practitioners have begun employing vanilla or fine-tuned LLMs alongside these established techniques to more effectively identify CWEs in code.~\cite{dontCWEAT} introduces CWEAT, a static analysis framework that parses RTL into abstract syntax trees (ASTs) and applies CWE-specific scanners to detect vulnerability-relevant code patterns. CWEAT focuses on security-relevant signals to narrow the search space during early design analysis. Self-HWDebug \cite{akyash2024selfhwdebugautomationllmselfinstructing} uses LLMs to automatically generate debugging instructions from example pairs of vulnerable and secure RTL corresponding to a specific CWE. These synthesized instructions are then reused to repair previously unseen RTL snippets within the same vulnerability class. The framework validates generated patches using assertion-based verification to ensure correctness. VerilogLAVD \cite{long2025veriloglavdllmaidedrulegeneration} represents RTL designs as property graphs combining structural and semantic dependencies. An LLM translates natural-language CWE descriptions into graph traversal rules, which are validated and executed to precisely localize vulnerabilities within the design. \cite{gadde2024artificialintelligencegenailens} proposes a data generation pipeline where LLMs produce large volumes of RTL samples targeting multiple hardware CWE classes. Formal verification techniques are then used to label each sample as vulnerable or CWE-free based on proof results and counterexamples, producing a large benchmark dataset for evaluating security-aware generative models. BugWhisperer~\cite{bugwhisperer} constructs a CWE-labeled RTL dataset from known SoC vulnerabilities and expands it using function-preserving variant generation. The authors then instruction-fine-tune open-source LLMs to classify whether RTL modules contain specific vulnerability types. LASHED~\cite{ahmad2025lashedllmsstatichardware} combines LLM reasoning with grounded static analysis. The LLM first identifies security-relevant assets such as keys, lock bits, and debug paths, after which the system generates targeted lint and formal verification checks. Finally, the LLM contextualizes the results to filter false positives and explain security risks; evaluation on four open-source SoCs across five CWEs shows high detection accuracy.

\subsection{Decoder-only transformer-based LLMs} Decoder-only transformer-based LLMs function as sequential text processors, transforming input text into contextual representations and predictive outputs. The process initiates with a tokenizer that converts the input text into a sequence of discrete tokens. Each token is subsequently mapped through an embedding layer into a continuous vector space of dimension ($d_{model}$), establishing the initial representation of the input.

The embeddings are processed through a stack of ($L$) transformer decoder blocks. Within each block, token representations are iteratively refined using masked self-attention, which restricts each token's attention to permitted positions as specified by an attention mask, typically enforcing a causal left-to-right dependency. A feed-forward network then further transforms these representations. As embeddings propagate through successive layers, they accumulate contextual information from preceding tokens.

The final output comprises context-aware embeddings for each token position. In standard generative applications, these embeddings are projected to vocabulary logits, transformed into probability distributions using softmax, and employed to sample or predict the subsequent token in the sequence.

\section{Methodology}\label{sec:methodology}
In this section, we present our framework \texttt{VeriCWEty}. \texttt{VeriCWEty} leverages vector embeddings generated by a Verilog fine-tuned decoder-only LLM to capture the syntax and semantics of different CWEs. The framework has three stages: the data generation phase, the embedding generation phase, and the training and evaluation phases.

\begin{figure}[!ht]
    \centering
    \includegraphics[scale=0.45]{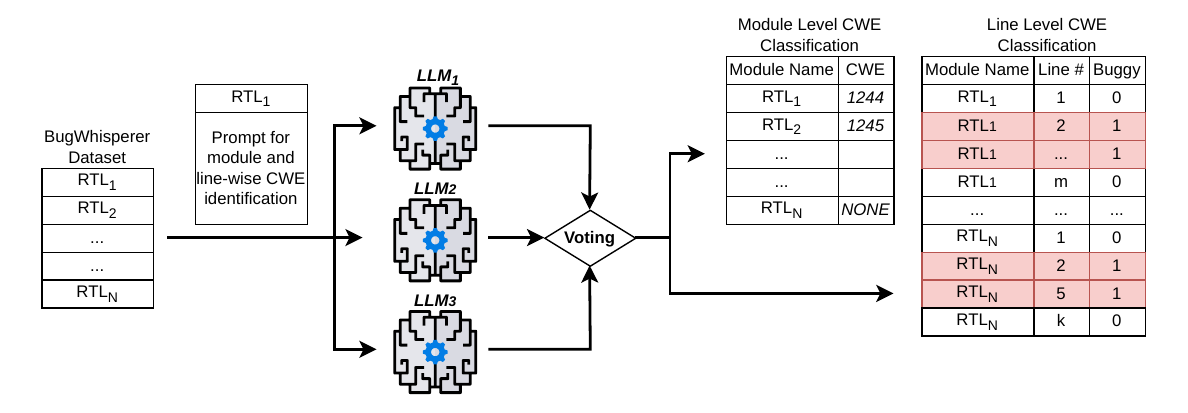}
    \caption{Voting scheme determines the module-level CWEs and line-level bugs}
    \label{fig:voting}
\end{figure}

\subsection{Dataset generation with voting models} We considered the open-source BugWhisperer dataset as our starting point. The dataset contains about 4000 buggy designs, but they are not labeled with the corresponding CWE. Since our aim is to capture the essence of the CWEs via vector embeddings, we did not resort to preprocessing steps such as AST generation or pattern matching~\cite{dontCWEAT} to identify key signals. Instead, we used a majority-rule approach to assign CWE labels to each data point. In this process, we used three LLMs: LLaMA 3.3-70B, GPT-4o-mini, and DeepSeek-V3.2. We also filtered out 1,000 bug-free datapoints from the Verigen dataset. We elaborate on the mechanism in Figure~\ref{fig:voting}. First, we use the Open Router~\cite{openrouter2026} setup to simultaneously query three different LLMs about the type of CWE present in the given datapoint. As per the BugWhisperer paper, we ask the LLMs to classify the designs into the following CWEs: 250, 269, 284, 310, 321, 506, 1198, 1244, 1245, 1260, and 1271. We maintain the label `NONE' for cases where the LLMs do not return any CWE. We also ask the LLMs to return the particular section/ lines of the code that they find buggy. Based on the response, a vote is taken, and the CWE reported by at least two models is selected. At the end of the data generation phase, we have two data sets. In one set, the design’s module name and the corresponding CWE are recorded. In the second set, each line of each module is labeled 1 or 0, indicating whether a bug is present or absent. 

\begin{figure*}[ht]
    \centering
    \includegraphics[scale=0.42]{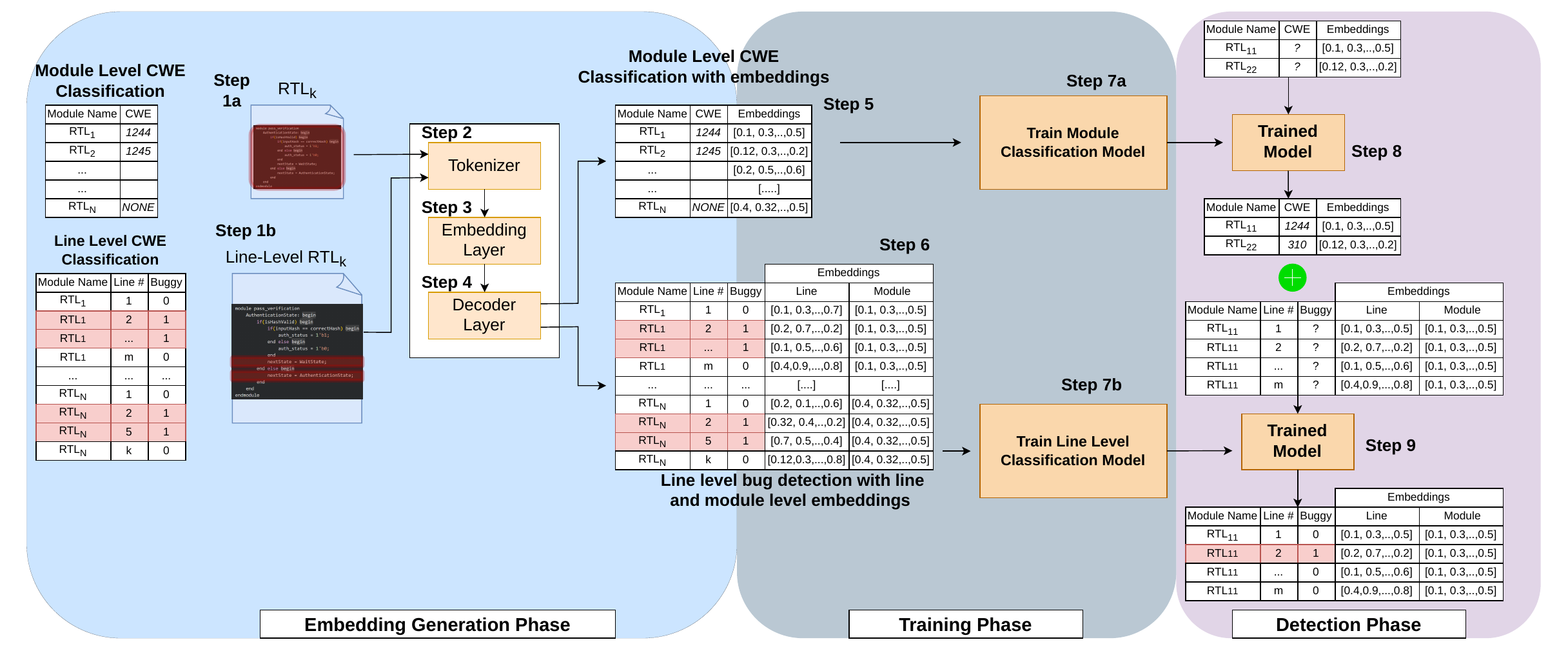}
    \caption{End-to-end \texttt{VeriCWEty} pipeline including data generation, embedding extraction, training, and inference}
    \label{fig:flow}
\end{figure*}

\subsection{Embedding generation Phase} Figure~\ref{fig:flow} illustrates the full pipeline of VeriCWEty. The two datasets obtained from the previous phase are handled separately (Refer to Step 1a and 1b in Figure~\ref{fig:flow}). Our framework follows the embedding retrieval framwork mentioned in TrojanLOC~\cite{xiao2025trojanloc}. To generate the module's embedding, the full design is sent as input to the tokenizer (Step 2). The tokenizers  break the text into tokens, which are then converted to vector embeddings (Step 3 and 4). For each module, there will be a vector embedding of dimension $d_{model}$ specific to the LLM model being used. Similarly, line-by-line embeddings are generated for each module. So, if a module has $n$ lines, $n \times d$-dimensional  vector embeddings will be generated. The module embeddings are combined with their corresponding CWE labels to generate the classifier's training data (Step 5). For the line-by-line scenario (Step 6), the embedding for each line is appended, along with the module embedding, to provide deeper context about the overall design and its relevance to the current line. These combined embeddings are then paired with the corresponding labels (1 or 0), forming the dataset for detecting line-level CWEs.

\begin{table}[t]
\centering
\small
\setlength{\tabcolsep}{4pt}
\caption{Performance with initial annotation (baseline) using weak models}
\renewcommand{\arraystretch}{1.1}
\begin{tabular}{lcccc}
\hline
\textbf{Class} & \textbf{Prec.} & \textbf{Rec.} & \textbf{F1} & \textbf{Sup.} \\
\hline
CWE-1244            & 0.547 & 0.506 & 0.526 & 81 \\
CWE-1245            & 0.431 & 0.383 & 0.405 & 81 \\
CWE-1260            & 0.529 & 0.450 & 0.487 & 20 \\
CWE-1271            & 0.000 & 0.000 & 0.000 & 3  \\
CWE-284             & 0.915 & 0.667 & 0.771 & 81 \\
CWE-310-AES-LEAK    & 0.822 & 0.914 & 0.866 & 81 \\
CWE-310-CSR-UNAUTH  & 0.625 & 0.741 & 0.678 & 81 \\
NONE                & 0.323 & 0.395 & 0.356 & 81 \\
\hline
\textbf{Accuracy}   &       &       & 0.591 & 509 \\
\textbf{Macro Avg}  & 0.524 & 0.507 & 0.511 & 509 \\
\textbf{Weighted Avg} & 0.604 & 0.591 & 0.592 & 509 \\
\hline
\end{tabular}
\label{tab:classification_results_vote}
\end{table}

\subsection{Training and Detection Phase} The module-level CWE-labeled dataset and the line-level buggy or non-buggy labeled dataset are divided using an 80-20\% split. A classifier is initially trained on the module-level dataset (Step 7), and the trained model is subsequently evaluated on the test set of buggy modules (Step 8). Predicted CWEs are assigned to each test module. For line-level testing, module-level testing is first conducted to identify the CWE type. Module-level embeddings are then combined with line embeddings for evaluation. The classifier predicts which lines within the designs are buggy.

\section{Results}
\subsection{Experimental Setup} The entire framework was implemented in Python 3.11.9. During the creation of the initial dataset, which consisted of 4000 CWE-annotated datapoints from the BugWhisperer and Verigen datasets, votes were collected from LLaMA 3.3-70B, GPT-4o-mini, and DeepSeek-V3.2. Embeddings were extracted using the open-source model `ajn313/cl-verilog-1.0'~\cite{clverilog}, which is based on Code LLaMA and LoRA-fine-tuned on the Verilog GitHub dataset, comprising approximately 13 billion parameters, and can be found on HuggingFace. For classification tasks, the XGBoostClassifier was employed. The classifier is implemented using an XGBoost model configured with a binary logistic objective, 300 estimators, a maximum tree depth of 6, and a learning rate of 0.1. To improve generalization, subsampling and column sampling rates are set to 0.8, with a minimum child weight of 3 and L2 regularization ($\lambda$ = 1.0). The model uses the histogram-based tree construction method for efficiency, incorporates class imbalance handling via `scale\_pos\_weight', optimizes using log-loss as the evaluation metric, and ensures reproducibility with a fixed random state while utilizing all available CPU cores (n\_jobs = -1).
The `cl-verilog-1.0' model was executed on an Nvidia A100 80 GB GPU to extract embeddings.

\subsection{CWE Annotation with weak closed source models} We prepared our initial dataset with LLaMA 3.3-70B, GPT-4o-mini, and DeepSeek-V3.2. 
On doing the module level classification, we observed very poor accuracy (0.591), precision and recall values. Table ~\ref{tab:classification_results_vote} highlights the f1-score, precision and recall of the various CWE classes. It was obvious that the problem was with the quality of the dataset. 

\begin{table}[t]
\centering
\caption{Number of mis-prediction of CWE by different models}
\label{tab:model_mismatch}
\begin{tabular}{lccc}
\toprule
\textbf{Model} & \textbf{Total Samples} & \textbf{Mismatches} & \textbf{Correct\%} \\
\midrule
Deepseek-v3.2 & 50 & 34 & 32 \\
LLaMA-3.3-70b-instruct & 50 & 36 & 28 \\
GPT-4o-mini & 50 & 25 & 50 \\
\hline
Gemini3 & 50 & 10 & 80 \\
GPT-5-Nano & 50 & 4 & 92 \\
GPT-5.4 & 50 & 3 & 94 \\
\bottomrule
\end{tabular}
\end{table}

\begin{table}[t]
\centering
\caption{Distribution of CWE Types in Training and Test Datasets}
\label{tab:cwe_distribution_train_test}
\begin{tabular}{lcccc}
\toprule
\textbf{CWE Type} & \textbf{\# Train} & \textbf{Train (\%)} & \textbf{\# Test} & \textbf{Test (\%)} \\
\midrule
CWE-1244 & 1106 & 38.05 & 276 & 37.96 \\
CWE-1245 & 779  & 26.80 & 213 & 29.30 \\
NONE     & 561  & 19.30 & 120 & 16.51 \\
CWE-310-AES-LEAK & 173 & 5.95 & 50  & 6.88 \\
CWE-321 & 103 & 3.54 & 26 & 3.58 \\
CWE-1271 & 55 & 1.89 & 16 & 2.20 \\
CWE-310-AES-DOS & 53 & 1.82 & 9  & 1.24 \\
CWE-310-CSR-UNAUTH & 30 & 1.03 & 5  & 0.69 \\
CWE-1260 & 28 & 0.96 & 7  & 0.96 \\
CWE-506 & 18 & 0.62 & 5  & 0.69 \\
CWE-1198 & 1 & 0.03 & 0  & 0.00 \\
\midrule
\textbf{Total} & \textbf{2907} & \textbf{100.00} & \textbf{727} & \textbf{100.00} \\
\bottomrule
\end{tabular}
\end{table}

{\flushleft \bf Observation.} LLaMA 3.3-70B and DeepSeek-v3.2 are not well trained on verilog code and also have limited exposure to CWEs. Thus they often returned incorrect labels. There were multiple occasions when both of them had the same wrong labels and outvoted GPT-4o-mini, which was comparatively giving more correct labels. An analysis of the mispredictions of these three models across 50 different designs with various CWEs is presented in Table~\ref{tab:model_mismatch}.
\begin{table}[t]
\centering
\small
\setlength{\tabcolsep}{4pt}
\caption{“Final performance after improved annotation and training using stronger latest models like GPT-5.4, GPT-5-Nano, and Gemini3 for labeling the dataset}
\renewcommand{\arraystretch}{1.1}
\begin{tabular}{lcccc}
\hline
\textbf{Class} & \textbf{Prec.} & \textbf{Rec.} & \textbf{F1} & \textbf{Sup.} \\
\hline
CWE-1244            & 0.886 & 0.870 & 0.878 & 276 \\
CWE-1245            & 0.884 & 0.789 & 0.834 & 213 \\
CWE-310-AES-DOS     & 0.556 & 0.556 & 0.556 & 9   \\
CWE-310-AES-LEAK    & 0.576 & 0.680 & 0.624 & 50  \\
CWE-310-CSR-UNAUTH  & 0.333 & 0.200 & 0.250 & 5   \\
CWE-321             & 0.477 & 0.808 & 0.600 & 26  \\
CWE-506             & 0.000 & 0.000 & 0.000 & 5   \\
NONE                & 0.555 & 0.592 & 0.573 & 120 \\
\hline
\textbf{Accuracy}   &       &       & 0.767 & 704 \\
\textbf{Macro Avg}  & 0.533 & 0.562 & 0.539 & 704 \\
\textbf{Weighted Avg} & 0.777 & 0.767 & 0.769 & 704 \\
\hline
\end{tabular}
\label{tab:classification_results_gpt5}
\end{table}

Also, the presence of generic CWEs, such as CWE-284 and CWE-269, contributed to the misclassifications. Classes like CWE-284 are indirectly related to CWE-1260, CWE-1244, and CWE-1245. Thus, on many occasions when the LLMs should have precisely identified CWE-1244 or CWE-1245, they misclassified it as CWE-284, which may not be incorrect but is very generic and is discouraged for bug classification on the CWE website.

\begin{figure*}
    \centering
    \includegraphics[width=1\linewidth]{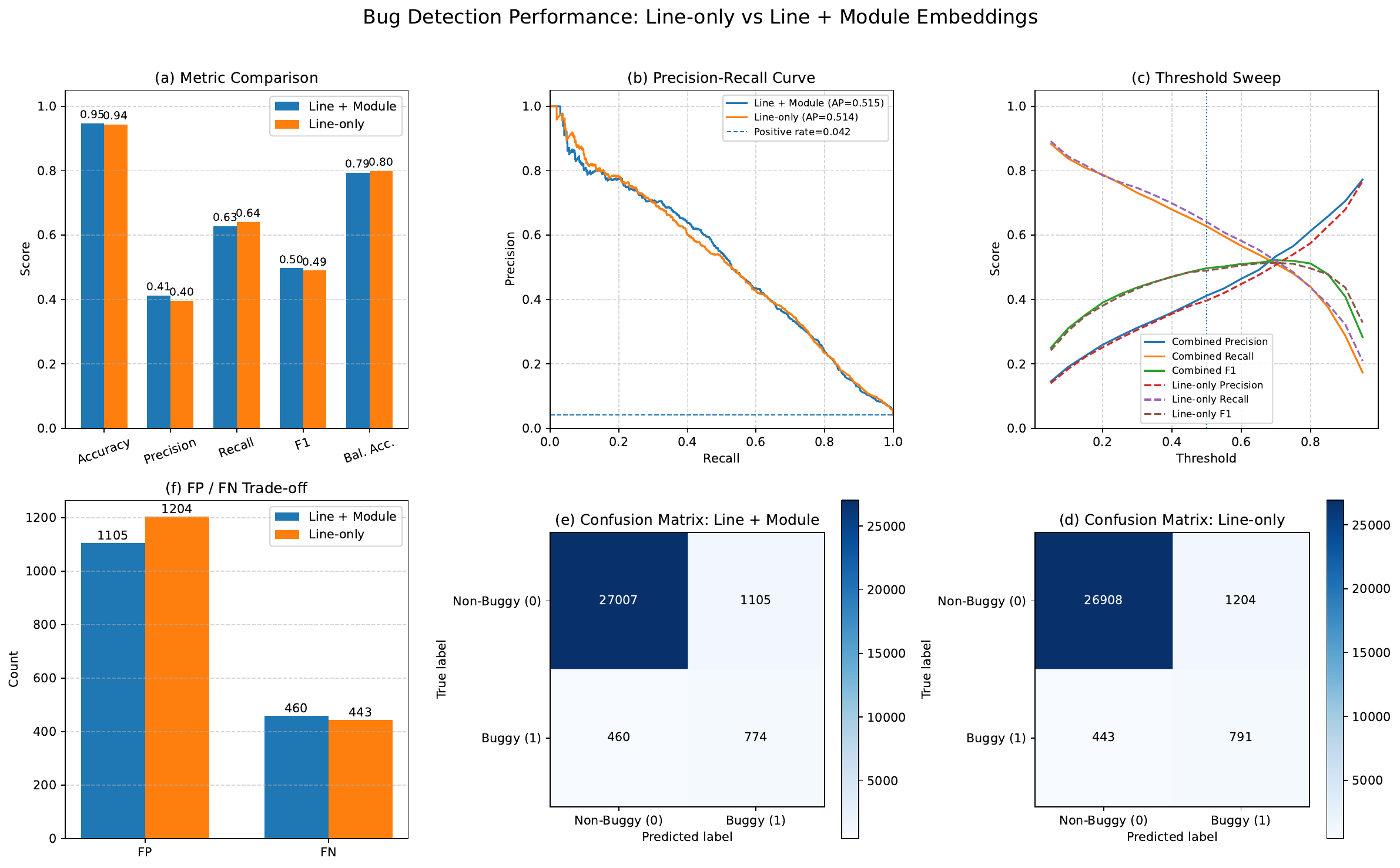}
    \caption{Line-level embeddings vs. Line-level + Module-level embeddings classification analysis. Starting from top left and going clock-wise (a) Metric Comparison (b) Precision-Recall Curve (c) Threshold Settings (d) Confusion Matrix for only Line-level embeddings for test (e) Confusion Matrix for Line-level combined with Module-level embeddings for test (f) FP/FN trade off}
    \label{fig:line_level}
\end{figure*}

{\flushleft \bf Solution.} We first improved our prompt to be more specific, avoiding generic CWEs such as CWE-284, CWE-250, and CWE-269. We tried to keep only those CWEs that are mostly independent of each other and specific to hardware (see Table~\ref{tab:cwe_distribution_train_test}).

We used three different and more recent models, Gemini3, GPT-5-Nano and GPT-5.4, to perform the voting. As shown in Table~\ref{tab:model_mismatch}, the precision of these models are significantly higher. Using them ensured the quality of the dataset, that we used for the next steps. 

\subsection{Module-level CWE classification} As discussed in Section~\ref{sec:methodology}, we used the open-source `ajn313/cl-verilog-1.0' to extract both the module-level as well as the combined module and line-level embeddings. The module-level embeddings and the corresponding CWE acted as the feature vector and the label. We split the dataset and into train and test data. Table~\ref{tab:classification_results_gpt5} shows that the precision, recall, and f1-score for CWE-1244 have gone up by 33\%, 36\%, and 35\% respectively. Similar trends can be seen for CWE-1245 and CWE-310-AES-DOS. 

\subsection{Line-level Bug Detection}
After the modules have been classified into their respective CWEs, the next step is to detect bugs at the line-level granularity. At first, we used only the line embeddings for each line of a given design module. From the 4000 datapoints of the initial dataset, we select 3000 unique modules and their corresponding line embeddings, totaling 148500 lines of code (LOC). We do an 80-20\% split on this dataset to prepare the train and test data, respectively. It is important to point out that the number of lines labeled as buggy (1) is significantly rare (6300 out of 148500 LOC) and accounts for only 4.2\% of the total dataset. 

Even with this significant imbalance, our XGBoost classifier performed quite well, with accuracy, precision, recall, f1-score, and balanced accuracy of 0.94, 0.40, 0.64, and 0.80, respectively. The confusion matrix at  Figure~\ref{fig:line_level}d shows that out of 1234 buggy lines, 791 lines were successfully detected as buggy, which is remarkable for such an imbalanced dataset.

To add more context to the overall design while classifying, we appended the module embedding to each line's embedding. During training and evaluation, we observed that adding the module embedding resulted in slight improvements in accuracy (0.95), precision (0.41), and f1-score (0.50). There was a negligible 1\% reduction in recall. As depicted in the plot Figure~\ref{fig:line_level}f, False Positive (FP) has reduced on the addition of the module embeddings. The Precision-Recall curve (Refer Figure~\ref{fig:line_level}b) shows that using only line embedding does not significantly affect the model's performance.

\subsection{Comparison with latest works} Self-HWDebug~\cite{akyash2024selfhwdebugautomationllmselfinstructing} uses a pair of correct and buggy code, and asks an LLM how to fix the buggy code. Based on the feedback provided, it asks another open-source LLM to fix the code. Always finding a buggy, non-buggy pair in the real world is not realistic. Also a same bug can be manifested in different ways. This technique will not be able to detect them. The authors have tested it on a limited number of 5-CWEs. \texttt{VeriCWEty} does not depend on a non-buggy version of the buggy code, and works on a dataset that has 11 distinct CWEs.

VerilogLAVD's ~\cite{long2025veriloglavd} knowledge assisted LLM approach shows the best results only on closed sourced 77 designs. Although they report 11 CWE's, there overall f1-score does not closs 0.57. For improper access control and finite state machine related bugs, \texttt{VeriCWEty} (f1-score 0.87 and 0.83) out performs VerilogLAVD (f1-score 0.70 and 0.67).

\section{Conclusion}
Conventionally, designers use methods such as formal, static, and simulation-based analysis to debug complex architectures. The scope of these techniques is often narrow, and their primary focus is on verifying the design's functional aspects rather than evaluating security vulnerabilities. To overcome these limitations, recent works such as BugWhisperer, VerilogLAVD, and Self-HWDebug have used LLMs, often combined with the established techniques mentioned above. Despite that, most CWE detections are limited to the module level, and detecting a wide variety of CWEs remains a challenge. Our framework, \texttt{VeriCWEty}, seeks to fill this gap by leveraging the power of vector embeddings generated as intermediate outputs in the LLM's flow. Vector embeddings generated by the decoder layer of a Verilog-finetuned transformer-based LLM can capture the syntactic and semantic nuances unique to each CWE. We leverage this property of embeddings and train a classifier on the module-level and line-level embeddings of various buggy designs. Our trained classifier can successfully detect bugs such as CWE-1244 and CWE-1245 with 89\% precision and pinpoint the lines containing the bugs with 96\% accuracy. Our work is the first to provide both modular-level and line-level granularity in CWE detection.

\bibliographystyle{ieeetr}
\bibliography{Ref}
\end{document}